\documentclass[%
 aip,
 amsmath,amssymb,
 reprint,%
]{revtex4-1}

\usepackage{graphicx}% Include figure files
\usepackage{dcolumn}% Align table columns on decimal point
\usepackage{bm}% bold math
\usepackage{siunitx}

\usepackage[utf8]{inputenc}
\usepackage[T1]{fontenc}
\usepackage{mathptmx}
\usepackage{etoolbox}

\usepackage{CJKutf8}
\makeatletter
\def\@email#1#2{%
 \endgroup
 \patchcmd{\titleblock@produce}
  {\frontmatter@RRAPformat}
  {\frontmatter@RRAPformat{\produce@RRAP{*#1\href{mailto:#2}{#2}}}\frontmatter@RRAPformat}
  {}{}
}%
\makeatother
\begin{document}
\begin{CJK*}{UTF8}{gbsn}

\preprint{AIP/123-QED}

\title{Effects of Flagellar Morphology on Swimming Performance and Directional Control in Microswimmers}
% Force line breaks with \\

\author{Baopi Liu(刘葆僻)}
\email{bpliu@csrc.ac.cn}
\email{bpliu@mail.bnu.edu.cn}
\affiliation{Complex Systems Division, Beijing Computational Science Research Center, Beijing 100193, China}

\author{Lu Chen(陈璐)}
\affiliation{College of Physics, Changchun Normal University, Changchun, Jilin 130032, China}

\author{Wenjun Xu(徐文君)}
\affiliation{School of Humanities and Basic Sciences, Shenzhen Institute of Information Technology, Shenzhen 518116, China}

\date{\today}% It is always \today, today,
             
\begin{abstract}
In a fluid environment, flagellated microswimmers propel themselves by rotating their flagella. The morphology of these flagella significantly influences forward speed, swimming efficiency, and directional stability, which are critical for their survival. This study begins by simulating the three-dimensional motion trajectories of microswimmers to analyze their kinematic characteristics. The simulation results demonstrate that microswimmers can actively adjust their forward direction by modifying the orientation of their flagella. We subsequently perform numerical simulations to visualize the flow fields generated by a microswimmer and examine the hydrodynamic interactions between the cell body and the flagella, focusing on their impacts on forward speed and swimming efficiency. We conclude that forward speed and swimming efficiency are closely related to the filament radius, pitch angle, and contour length of the flagella, while the yaw angle of locomotion is determined by the helix radius and contour length of the flagella. We conclude that the pitch angle for maximum forward speed is slightly smaller than that for maximum swimming efficiency, which suggests that microswimmers can effectively alternate between states of maximum forward speed and maximum swimming efficiency by fine-tuning their pitch angle and adapting to varying ecological conditions. These morphological characteristics of microswimmers may result from species competition and natural selection. This research establishes an optimized model for microswimmers, providing valuable insights for the design of enhanced microrobots tailored to specific applications.
\end{abstract}
\maketitle
\end{CJK*}
%%%%%%%%%%%%%%%%%%%%%%%%%%%%%%%%%%%%%%%%%%%%%%%%%%%%%%%%%%%%%%%%%%%%%%%%%%%%%%%%%%%%%%%%%%%%%%%%%%%%%%
\section{Introduction}
%%%%%%%%%%%%%%%%%%%%%%%%%%%%%%%%%%%%%%%%%%%%%%%%%%%%%%%%%%%%%%%%%%%%%%%%%%%%%%%%%%%%%%%%%%%%%%%%%%%%%%
Flagellated bacteria are a crucial group of microorganisms playing a vital role in natural ecosystems primarily due to their unique swimming mechanisms and exceptional motility. These bacteria are typically composed of a cell body, flagella, hooks, and motors, with the beating or rotation of the flagella serving as the primary mechanism for self-propulsion~\cite{Silverman1974,Lauga2009,Lauga2016}. Research on bacterial motility has not only advanced the fields of microbiology and physics, but also provided significant insights into ecological functions~\cite{Guasto2012}. The morphological characteristics of these bacteria have evolved over time to adapt to diverse living environments. This evolution directly influences their survival strategies and enhances their motility through their specific morphological structures~\cite{Roszak1987,Young2006,Mitchell2006,Justice2008,Spagnolie2011,Van2017}.

Marine bacteria exhibit rapid forward speeds, high swimming efficiencies, and strong directional responses to their ecological environment. These traits enable them to quickly react to chemical gradients, ensuring efficient nutrient acquisition and reproduction~\cite{Maki2000,Stocker2012}. Previous studies have shown that the morphological characteristics of flagella, such as filament radius, pitch angle, and contour length, significantly influence bacterial forward speed, swimming efficiency, and motility stability~\cite{Spagnolie2011,Purcell1997,Li2006,Chattopadhyay2006,Najafi2018,Nguyen2018}. Collectively, these factors determine the competitiveness of bacteria in complex environments and thereby influence their survival and reproduction~\cite{Fauci2006}.

In recent years, rapid advancements in experimental techniques and numerical simulation capabilities have enabled researchers to explore the physical mechanisms of bacterial motility in greater depth~\cite{Silverman1974,Bianchi2017,Thawani2018,Zottl2019,Colin2021}. The locomotion of bacteria in fluids generates specific flow fields that not only influence the swimming trajectories of microswimmers but also significantly affect the movement of surrounding microorganisms. Understanding these flow characteristics is essential for a comprehensive understanding of the motion mechanisms of microorganisms and the hydrodynamic interactions among their components~\cite{Zottl2019,Drescher2010,Drescher2011,Mathijssen2015}. Variations in parameters such as flagellar filament radius, helix radius, pitch angle, and axial length can significantly impact the forward speed, swimming efficiency, precession angle (also referred to as the wobbling angle), and yaw angle of bacteria~\cite{Shum2010,Son2016,Mousavi2020,Tang2020f,Tokarova2021}. Therefore, establishing an appropriate model of single-flagellated microswimmers and conducting systematic simulation studies are crucial for revealing their optimal morphological characteristics.

In-depth research on the motion mechanisms of microswimmers provides valuable insights into their biological characteristics and serves as a key source of inspiration for the design of new microrobots. With the rapid advancements in microbiology and engineering technologies, researchers have conducted more comprehensive investigations into the motion mechanisms and optimal morphologies of single-flagellated microswimmers~\cite{Nelson2010,Temel2015,Ng2021}. Microrobots, as an emerging technology, demonstrate wide-ranging application potential across various fields, including biomedicine, environmental monitoring, and micro-manipulation~\cite{Palagi2018,Zhou2021}. By harnessing the motion mechanisms of single-flagellated bacteria, researchers can design more efficient and flexible microrobots that can adapt to complex working environments~\cite{Huang2019}.

In this study, we systematically investigate the dynamic and kinematic characteristics of a helical flagellated microswimmer model in a viscous fluid using numerical simulation methods. By employing the Twin Multipole Moment (TMM) method to calculate the resistance matrix of the system, we effectively account for the hydrodynamic interactions between the cell body and the flagella. The TMM method considers both near-field and far-field interactions, making it suitable for high-resolution studies of the hydrodynamics of microswimmers~\cite{Jeffrey1984,Liu2025}. We adjust several morphological parameters of the microswimmer model, including the filament radius, helix radius, pitch angle, and contour length of the flagella, to assess its motion trajectory, flow field, forward speed, swimming efficiency, and yaw angle. These simulations help us determine the optimal parameters for the microswimmer design. Collectively, these factors influence the competitiveness of bacteria in complex environments, which subsequently impacts their survival and reproductive success in various contexts.

The structure of this paper is organized as follows: In Sec.~\uppercase\expandafter{\romannumeral2}, we establish the microswimmer simulation model and analyze its kinematics and swimming efficiency. In Sec.~\uppercase\expandafter{\romannumeral3. A}, we simulate the trajectory of the microswimmer and analyze its kinematic characteristics. Following that, in Sec.~\uppercase\expandafter{\romannumeral3. B}, we examine the flow field generated by the microswimmer during swimming, specifically analyzing how the hydrodynamic interactions between the cell body and the flagella affects its forward speed and swimming efficiency. In Sec.~\uppercase\expandafter{\romannumeral3. C}, we conduct a detailed numerical simulation to assess the impact of varying filament radius, pitch angle, and contour length on the forward speed and swimming efficiency of the microswimmer, ultimately identifying the optimal filament radius, pitch angle, and contour length. Additionally, we analyze how variations in contour length and helix radius influence forward speed and swimming efficiency, integrating the yaw angle and helical diameter into our examination of locomotion trajectories at different helical radii to identify the optimal helix radius. Finally, in Sec.~\uppercase\expandafter{\romannumeral4}, we summarize the main findings of our investigation.

%%%%%%%%%%%%%%%%%%%%%%%%%%%%%%%%%%%%%%%%%%%%%%%%%%%%%%%%%%%%%%%%%%%%%%%%%%%%%%%%%%%%%%%%%%%%%%%%%%%%%%
\section{Methodology}
%%%%%%%%%%%%%%%%%%%%%%%%%%%%%%%%%%%%%%%%%%%%%%%%%%%%%%%%%%%%%%%%%%%%%%%%%%%%%%%%%%%%%%%%%%%%%%%%%%%%%%
\subsection{Model and Methods}
%%%%%%%%%%%%%%%%%%%%%%%%%%%%%%%%%%%%%%%%%%%%%%%%%%%%%%%%%%%%%%%%%%%%%%%%%%%%%%%%%%%%%%%%%%%%%%%%%%%%%%
When microswimmers move through a viscous medium, the viscous forces they experience significantly exceed the inertial forces. Consequently, the fluid dynamics of the microswimmers can be effectively described using the incompressible Stokes equations~\cite{Happel2012,Kim2013}:
\begin{equation}
\begin{split}
&\mu\nabla^{2}\mathbf{u}-\nabla p=-\mathbf{f},\\
&\nabla\cdot\mathbf{u}=0
\label{eq:refname1}
\end{split}
\end{equation}
where $\mu$ is the dynamic viscosity, $\mathbf{u}$ represents the fluid velocity, $p$ signifies the pressure, and $\mathbf{f}$ corresponds to the external force density.

The linearity of the Stokes equations implies a linear relationship between the forces $\mathbf{F}$ and torques $\mathbf{T}$ exerted by the microswimmer on the fluid, and its translational velocities $\mathbf{U}$ and rotational velocities $\boldsymbol{\Omega}$:
\begin{equation}
\left(\begin{matrix}\mathbf{F}\\ \mathbf{T}\end{matrix}\right)=\mathcal{R}\left(\begin{matrix}\mathbf{U}-\mathbf{U}^{\infty}\\ \boldsymbol{\Omega}-\boldsymbol{\Omega}^{\infty}\end{matrix}\right).
\label{eq:refname2}
\end{equation}
where $\mathbf{U}^{\infty}$ and $\boldsymbol{\Omega}^{\infty}$ denote the ambient flow fields, while $\mathbf{U}-\mathbf{U}^{\infty}$ and $\boldsymbol{\Omega}-\boldsymbol{\Omega}^{\infty}$ represent the relative velocities of the $N$ spheres in a $3N$-dimensional space relative to the ambient flow. The resistance matrix of the system, denoted as $\mathcal{R}$, can be calculated using TMM~\cite{Liu2025}. The reciprocal relation that relates the velocities of the microswimmer to the external forces and torques in the ambient flow fields is expressed using the mobility matrix $\mathcal{M}$, which can be expressed as
\begin{equation}
\left(\begin{matrix}\mathbf{U}\\ \boldsymbol{\Omega}\end{matrix}\right)=\mathcal{M}\left(\begin{matrix}\mathbf{F}\\ \mathbf{T}\end{matrix}\right)+\left(\begin{matrix}\mathbf{U}^{\infty}\\ \boldsymbol{\Omega}^{\infty}\end{matrix}\right).
\label{eq:refname3}
\end{equation}

\begin{figure}
\centering
\includegraphics[width=0.50\textwidth]{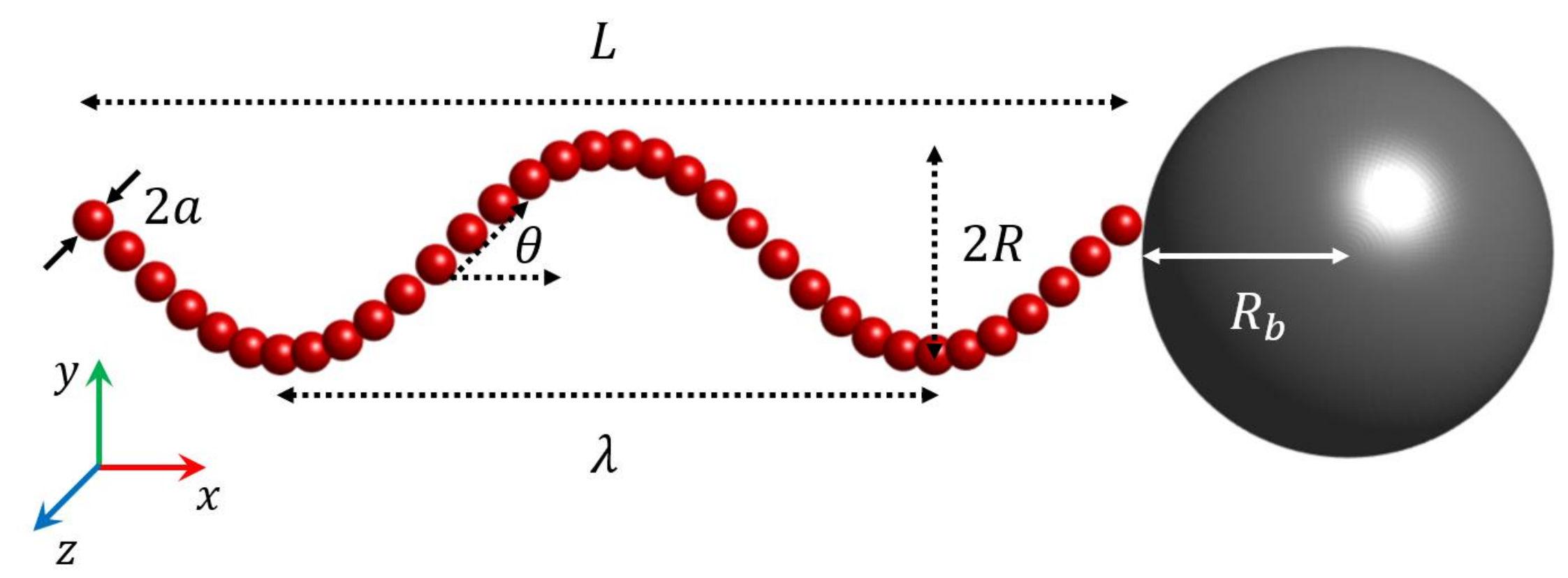}
\caption{Schematic representation of a microswimmer model featuring a spherical cell body and a helical flagellum. The cell body has a radius $R_{b}$, while the flagellum is characterized by its helix radius $R$, filament radius $a$, pitch $\lambda$, pitch angle $\theta$, axial length $L$, and contour length $\Lambda=L/\cos\theta$, where $\tan\theta=2\pi R/\lambda$.}
\label{fig:fig1}
\end{figure}

Flagellated microswimmers generally consist of a cell body, flagella, motors, and hooks~\cite{Silverman1974,Lauga2009,Lauga2016,Nord2022}. In most numerical simulations of rigid microswimmers, the effects of hooks are typically neglected~\cite{Abbott2009,Acemoglu2014,Vizsnyiczai2020}. In this study, we assume that the specific morphology of the hooks can be neglected and choose to omit them. The microswimmer is modeled as a rigid helical flagellum attached to a spherical cell body with a radius $R_{b}$, as illustrated in Fig.~\ref{fig:fig1}. The helical flagellum is characterized by its helix radius $R$, pitch $\lambda$, contour length $\Lambda$, and filament radius $a$. Its axial length is given by $L=\Lambda\cos\theta$, where the pitch angle $\theta$ satisfies $\tan\theta=2\pi R/\lambda$. The center of the cell body is denoted by $\mathbf{r}_{b}$, while the coordinate of the connection point between the flagellar axis and the cell body is represented as $\mathbf{r}_{0}$. The associated material frame at the connection point $\mathbf{r}_{0}$ is defined as $\{\mathbf{D}_{1},\mathbf{D}_{2},\mathbf{D}_{3}\}$, where $\mathbf{D}_{1}$ is the unit vector along the flagellar axis pointing towards the cell body. We define the unit vector indicating the rotation direction of the microswimmer's motor as $\mathbf{e}_{m}=\frac{\mathbf{r}_{0}-\mathbf{r}_{b}}{|\mathbf{r}_{0}-\mathbf{r}_{b}|}$. In the laboratory coordinate system, the centerline positions of the left-handed helical flagellum are given by:
\begin{equation}
\begin{split}
\mathbf{r}=l\cos\theta\mathbf{D}_{1}+R\sin\phi\mathbf{D}_{2}+R\cos\phi\mathbf{D}_{3}+\mathbf{r}_{0}.
\label{eq:refname4}
\end{split}
\end{equation}
where $\phi=kl\cos\theta$ is the initial phase, and $k=2\pi/\lambda$ denotes the wave number. Here, $l\in[0,\Lambda]$ is the arc length of the flagellum, measured from the point $\mathbf{r}_{0}$.

The instantaneous translational and rotational velocities of the center of the cell body are denoted as $\mathbf{U}_{b}$ and $\boldsymbol{\Omega}_{b}$, respectively. The instantaneous translational and rotational velocities of any point along the flagellum are given by:
\begin{equation}
\begin{split}
&\boldsymbol{\Omega}_{t}(\mathbf{r})=\boldsymbol{\Omega}_{b}+\boldsymbol{\Omega}_{m}\\
&\mathbf{U}_{t}(\mathbf{r})=\mathbf{U}_{b}+\boldsymbol{\Omega}_{t}(\mathbf{r})\times\left(\mathbf{r}-\mathbf{r}_{b}\right).
\label{eq:refname5}
\end{split}
\end{equation}
where $\boldsymbol{\Omega}_{m}=2\pi f\mathbf{e}_{m}$ (with $f$ representing the motor's rotational frequency) denotes the angular velocity of the motor, and $\mathbf{r}$ is the position vector of the point along the flagellum. Since a freely moving microswimmer experiences no external forces or torques, we apply the conditions of force and torque balance to this microswimmer model
\begin{equation}
\begin{split}
&\mathbf{F}_{b}+\sum_{i=1}^{N-1}\mathbf{F}_{t}^{i}=0,\\
&\mathbf{T}_{b}+\sum_{i=1}^{N-1}\mathbf{T}_{t}^{i}+\sum_{i=1}^{N-1}\left(\mathbf{r}^{i}-\mathbf{r}_{b}\right)\times\mathbf{F}_{t}^{i}=0.
\label{eq:refname6}
\end{split}
\end{equation}
where $\mathbf{F}_{b}$ and $\mathbf{T}_{b}$ represent the force and torque exerted by the cell body on the fluid, respectively. Similarly, $\mathbf{F}_{t}$ and $\mathbf{T}_{t}$ denote the forces and torques exerted by the flagellum on the fluid. The number of spheres on the flagellum is $N-1$, while the total number of spheres in the microswimmer model is $N$.

%%%%%%%%%%%%%%%%%%%%%%%%%%%%%%%%%%%%%%%%%%%%%%%%%%%%%%%%%%%%%%%%%%%%%%%%%%%%%%%%%%%%%%%%%%%%%%%%%%%%%%
\subsection{Kinematic Analysis}
%%%%%%%%%%%%%%%%%%%%%%%%%%%%%%%%%%%%%%%%%%%%%%%%%%%%%%%%%%%%%%%%%%%%%%%%%%%%%%%%%%%%%%%%%%%%%%%%%%%%%%
Analyzing the kinematics of microswimmers is crucial for assessing their performance. The rigid structure of both the cell body and the flagellum causes non-axisymmetric behavior during rotation, resulting in precessional motion and a helical trajectory. The forward direction of the microswimmer is defined by the rotation of the flagellum, expressed as $\mathbf{e}_{t}=\boldsymbol{\Omega}_{t}/|\boldsymbol{\Omega}_{t}|$~\cite{Constantino2016,Benhal2021,Kamdar2022}. The forward speed $U_{f}$ is given by $U_{f}=-\mathbf{U}_{b}\cdot\mathbf{e}_{t}$, while the speed perpendicular to the forward direction is $U_{v}=\sqrt{U_{b}^{2}-U_{f}^{2}}$, where $\mathbf{U}_{b}$ is the translational velocity of the cell body. Consequently, the pitch of the helical trajectory is given by $H=\frac{2\pi}{|\boldsymbol{\Omega}_{t}|}U_{f}$, and the helical diameter is given by $D=\frac{2}{|\boldsymbol{\Omega}_{t}|}U_{v}$.

The angle between the flagellar axis and the direction of motor rotation is defined as $\psi=\arccos(-\mathbf{e}_{m}\cdot\mathbf{e}_{a})$, where $\mathbf{e}_{a}=\mathbf{D}_{1}$. The precession angle is given by $\phi=\arccos(-\mathbf{e}_{t}\cdot\mathbf{e}_{b})$, with $\mathbf{e}_{b}=\boldsymbol{\Omega}_{b}/|\boldsymbol{\Omega}_{b}|$ representing the spin direction of the cell body. The yaw angle, which quantifies the deviation of the microswimmer’s forward direction from the flagellar axis, is defined as $\beta=\arccos(-\mathbf{e}_{t}\cdot\mathbf{e}_{a})$. A smaller yaw angle indicates greater directional stability, as the axial direction of the flagellum is known, while the forward direction remains uncertain.

%%%%%%%%%%%%%%%%%%%%%%%%%%%%%%%%%%%%%%%%%%%%%%%%%%%%%%%%%%%%%%%%%%%%%%%%%%%%%%%%%%%%%%%%%%%%%%%%%%%%%%
\subsection{Swimming Efficiency}
%%%%%%%%%%%%%%%%%%%%%%%%%%%%%%%%%%%%%%%%%%%%%%%%%%%%%%%%%%%%%%%%%%%%%%%%%%%%%%%%%%%%%%%%%%%%%%%%%%%%%%
Swimming efficiency is an important parameter that characterizes the performance of a microswimmer. The propulsive force acting in the forward direction is expressed as $F_{f}=-\mathbf{F}_{b}\cdot\mathbf{e}_{t}$. The propulsive efficiency $\epsilon$ is defined as the ratio of the propulsive power output to the rotary power input provided by the motor, which refers to the power expended to rotate the flagellum~\cite{Purcell1997,Chattopadhyay2006,Spagnolie2011,Acemoglu2014}:
\begin{equation}
\begin{split}
\epsilon\equiv\frac{F_{f}\cdot U_{f}}{|\mathbf{T}_{b}\cdot\boldsymbol{\Omega}_{m}|}.
\label{eq:refname7}
\end{split}
\end{equation}
where $\mathbf{T}_{b}$ represents the torque exerted on the cell body by the motor, and $\boldsymbol{\Omega}_{m}$ denotes the rotational velocity of the motor. The various parameters of the model for the microswimmer are summarized in Table.~\ref{tab:table 1}.

\begin{table}
\caption{\label{tab:table 1}The various parameters in the numerical model and their values.}
\begin{ruledtabular}
\begin{tabular}{lp{4.3cm}l}
\textbf{Notation} & \textbf{Description} & \textbf{Value}\\
\hline
$\mu$ & Dynamic viscosity & $1.0$ \si{\mu g/(\mu m\cdot s)}\\
$R_{b}$ & Radius of cell body & $1.0$ \si{\mu m}\\
$f$ & Rotation frequency of motor & $100$ \si{Hz}\\
$\mathbf{e}_{a}$ & Flagellar axis direction & $\mathbf{e}_{a}=\mathbf{D}_{1}$\\
$\mathbf{e}_{m}$ & Rotation direction of motor & $\mathbf{e}_{m}=\frac{\mathbf{r}_{0}-\mathbf{r}_{b}}{|\mathbf{r}_{0}-\mathbf{r}_{b}|}$\\
$\boldsymbol{\Omega}_{m}$ & Angular velocity of motor & $2\pi f\mathbf{e}_{m}$\\
$\boldsymbol{\Omega}_{b}$ & Angular velocity of cell body & Variable\\
$\boldsymbol{\Omega}_{t}$ & Angular velocity of flagellum & $\boldsymbol{\Omega}_{t}=\boldsymbol{\Omega}_{b}+\boldsymbol{\Omega}_{m}$\\
$-\mathbf{e}_{t}$ & Forward direction & $-\boldsymbol{\Omega}_{t}/|\boldsymbol{\Omega}_{t}|$\\
$\psi$ & Angle between flagellar axis and rotation direction of motor & $\arccos(-\mathbf{e}_{m}\cdot\mathbf{e}_{a})$\\
$\beta$ & Yaw angle & $\arccos(-\mathbf{e}_{t}\cdot\mathbf{e}_{a})$\\
$U_{f}$ & Forward speed & $-\mathbf{U}_{b}\cdot\mathbf{e}_{t}$\\
$U_{v}$ & Speed perpendicular to forward direction & $\sqrt{U_{b}^{2}-U_{f}^{2}}$\\
$H$ & Pitch of helical trajectory & $\frac{2\pi}{|\boldsymbol{\Omega}_{t}|}U_{f}$\\
$D$ & Helical diameter of trajectory & $\frac{2}{|\boldsymbol{\Omega}_{t}|}U_{v}$\\
\end{tabular}
\end{ruledtabular}
\end{table}

%%%%%%%%%%%%%%%%%%%%%%%%%%%%%%%%%%%%%%%%%%%%%%%%%%%%%%%%%%%%%%%%%%%%%%%%%%%%%%%%%%%%%%%%%%%%%%%%%%%%%%
\section{Results}
%%%%%%%%%%%%%%%%%%%%%%%%%%%%%%%%%%%%%%%%%%%%%%%%%%%%%%%%%%%%%%%%%%%%%%%%%%%%%%%%%%%%%%%%%%%%%%%%%%%%%%
\subsection{Trajectories of Microswimmers}
%%%%%%%%%%%%%%%%%%%%%%%%%%%%%%%%%%%%%%%%%%%%%%%%%%%%%%%%%%%%%%%%%%%%%%%%%%%%%%%%%%%%%%%%%%%%%%%%%%%%%%
To determine the trajectory of the microswimmer, we utilize a stationary Cartesian coordinate system in the laboratory frame. The trajectory of the center of the cell body serves as the representative path of the microswimmer, and its coordinates are updated over time as follows:
\begin{equation}
\begin{split}
&\mathbf{r}_{b}(t+\Delta t)=\mathbf{r}_{b}(t)+\mathbf{U}_{b}(t)\Delta t.
\label{eq:refname8}
\end{split}
\end{equation}
Accordingly, the connecting point $\mathbf{r}_{0}$ is updated over time as follows:
\begin{equation}
\begin{split}
&\mathbf{r}_{0}(t+\Delta t)=\mathbf{r}_{b}(t+\Delta t)+R_{b}\mathbf{e}_{m}(t+\Delta t).
\label{eq:refname9}
\end{split}
\end{equation}
Here, the unit vector representing the direction of the motor's rotation, $\mathbf{e}_{m}$, evolves over time according to
\begin{equation}
\begin{split}
&\mathbf{e}_{m}(t+\Delta t)=\mathbf{R}(\mathbf{e}_{b},\theta_{b})\cdot\mathbf{e}_{m}(t).
\label{eq:refname10}
\end{split}
\end{equation}
In this equation, $\mathbf{e}_{b}=\boldsymbol{\Omega}_{b}(t)/|\boldsymbol{\Omega}_{b}(t)|$ and $\theta_{b}=|\boldsymbol{\Omega}_{b}(t)|\Delta t$ are instantaneous variables. The material frame corresponding to the connecting point is updated as follows:
\begin{equation}
\begin{split}
&\{\mathbf{D}_{1}(t+\Delta t),\mathbf{D}_{2}(t+\Delta t),\mathbf{D}_{3}(t+\Delta t)\}\\
=&\mathbf{R}(\mathbf{e}_{b},\theta_{b})\cdot\mathbf{R}(\mathbf{e}_{m},\theta_{m})\cdot\{\mathbf{D}_{1}(t),\mathbf{D}_{2}(t),\mathbf{D}_{3}(t)\}.
\label{eq:refname11}
\end{split}
\end{equation}
Here, $\mathbf{e}_{m}=\boldsymbol{\Omega}_{m}(t)/|\boldsymbol{\Omega}_{m}(t)|$ and $\theta_{m}=|\boldsymbol{\Omega}_{m}(t)|\Delta t$ are instantaneous variables. The centerline positions of the flagellum, as described by Eq.~\ref{eq:refname4}, are then updated as follows:
\begin{equation}
\begin{split}
\mathbf{r}(t+\Delta t)&=l\cos\theta\mathbf{D}_{1}(t+\Delta t)+R\sin\phi\mathbf{D}_{2}(t+\Delta t)\\ &+R\cos\phi\mathbf{D}_{3}(t+\Delta t)+\mathbf{r}_{0}(t+\Delta t).
\label{eq:refname12}
\end{split}
\end{equation}
The rotation matrix $\mathbf{R}(\mathbf{e},\theta)$, known as the Rodrigues' rotation matrix, represents a rotation through an angle $\theta=\vert\boldsymbol{\Omega}\vert\Delta t$ about the axis defined by the unit vector $\mathbf{e}=\boldsymbol{\Omega}/\vert\boldsymbol{\Omega}\vert$. The Rodrigues' rotation matrix is expressed using the following formula:~\cite{Dai2015,Murray2017}
\begin{equation}
\begin{split}
&\mathbf{R}(\mathbf{e},\theta)=\cos\theta\mathbb{I}+(1-\cos\theta)\mathbf{e}\otimes\mathbf{e}^{T}+\sin\theta\mathbf{e}\times.
\label{eq:refname13}
\end{split}
\end{equation}
Here, $\mathbb{I}$ denotes the $3\times3$ identity matrix, $\otimes$ represents the Kronecker product, and $(\mathbf{e}\times)$ is a $3\times3$ antisymmetric matrix defined such that $(\mathbf{e}\times)\mathbf{v}=\mathbf{e}\times\mathbf{v}$ for any vector $\mathbf{v}$.

\begin{figure}
\centering
\includegraphics[width=0.50\textwidth]{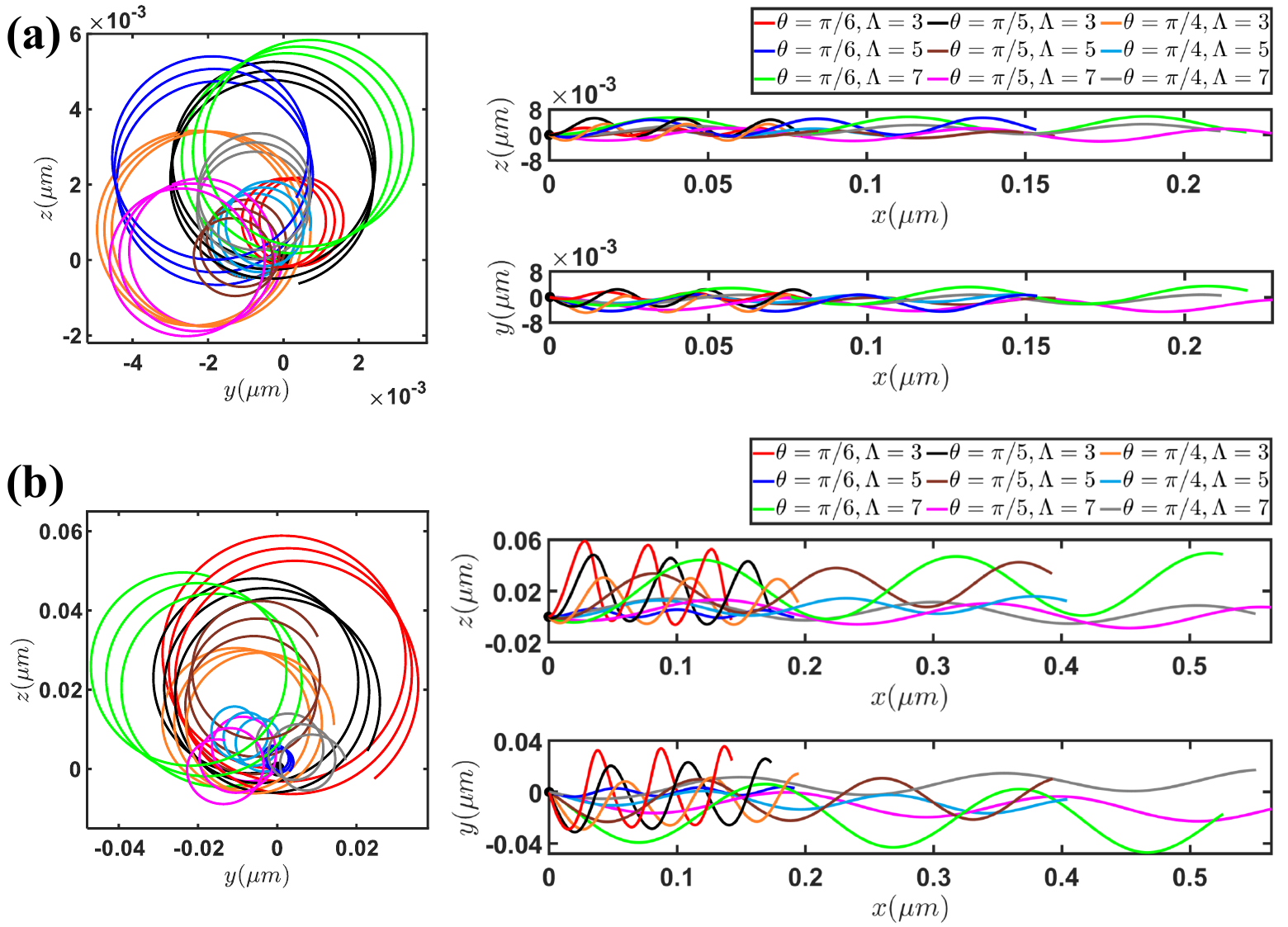}
\caption{(a) Trajectories of a freely swimming microswimmer with varying contour lengths and pitch angles at a helix radius of $R=0.2$ \si{\mu m}. (b) Trajectories of a freely swimming microswimmer with different contour lengths and pitch angles at a helix radius of $R=0.5$ \si{\mu m}.}
\label{fig:fig2}
\end{figure}

Figure~\ref{fig:fig2} illustrates the trajectories of microswimmers with varying contour lengths, pitch angles, and helix radii over a duration of $0.03$ \si{s}. It is clear that as the contour length increases, the forward speed of the microswimmer also increases. Furthermore, an increase in the helix radius $R$ corresponds to a larger helical diameter $D$ of the trajectories. Notably, when the pitch angle is $\pi/5$, the forward speed exceeds that of the pitch angles $\pi/6$ and $\pi/4$. The microswimmer's trajectory forms a cylindrical helix, with the forward direction $-\mathbf{e}_{t}$ is a fixed unit vector. From Fig.~\ref{fig:fig2}, it is apparent that the morphological parameters of different microswimmers lead to variations in forward speeds and helical radii of the trajectories, indicating the presence of an optimal flagellar morphology that enables the microswimmer to achieve maximum forward speed.

\begin{figure}
\centering
\includegraphics[width=0.50\textwidth]{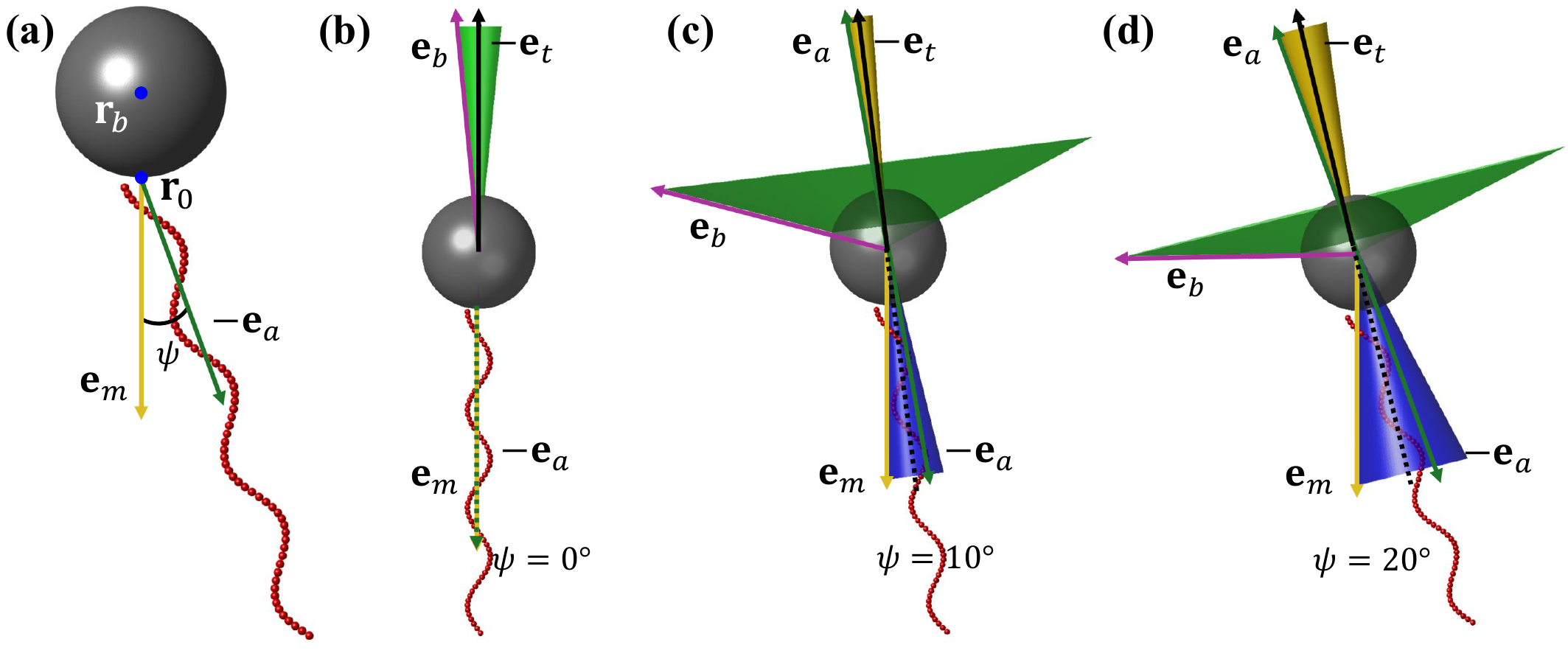}
\caption{(a) Schematic diagram of the microswimmer in its initial state. (b), (c), and (d) illustrate the cones formed by rotation in different directions around the forward axis over time, corresponding to various angles $\psi$ between the motor's rotation direction and the flagellar axis.}
\label{fig:fig3}
\end{figure}

When a microswimmer moves through a fluid, several specific directional vectors must be carefully considered. Fig.~\ref{fig:fig3} illustrates the time-dependent changes in these unit vectors during the locomotion of the microswimmers. In Fig.~\ref{fig:fig3}(a), the initial model of the microswimmer is depicted, with $\mathbf{e}_{m}$ representing the direction of the motor's rotation and $\mathbf{e}_{a}$ indicating the direction of the flagellar axis, separated by an angle $\psi$. Fig.~\ref{fig:fig3}(b) shows the variations of these directions over time when $\psi=0^{\circ}$. In this case, the forward direction remains unchanged, while the purple arrow denotes the spin direction of the cell body, which rotates around the forward axis to form a green cone. The angle between the spin direction and the forward direction is referred to as the precession angle, while the angle between the flagellar axis and the forward direction approaches zero, indicating that these two directions are nearly aligned. Fig.~\ref{fig:fig3}(c) illustrates the directional variations when $\psi=10^{\circ}$. In this scenario, the green cone represents the spin direction of the cell body rotating around the forward direction, the dark blue cone indicates the motor's rotation direction, and the gold cone denotes the flagellar axis rotating around the forward direction. Similarly, Fig.~\ref{fig:fig3}(d) presents the variations of these directions when $\psi=20^{\circ}$.

As shown in Figs.~\ref{fig:fig3}(b)-(d), the precession angle is the largest, followed by the angle between the motor's rotation and the forward direction. Conversely, the angle between the flagellar axis direction and the forward direction is the smallest. When the motor's rotation direction aligns with the flagellar axis, the angle between the forward direction and the flagellar axis approaches a minimum, which we define as the yaw angle. Since the direction of the flagellum is controllable, the forward direction of the microswimmer can be adjusted by manipulating the orientation of the flagellum.

\begin{figure}
\centering
\includegraphics[width=0.50\textwidth]{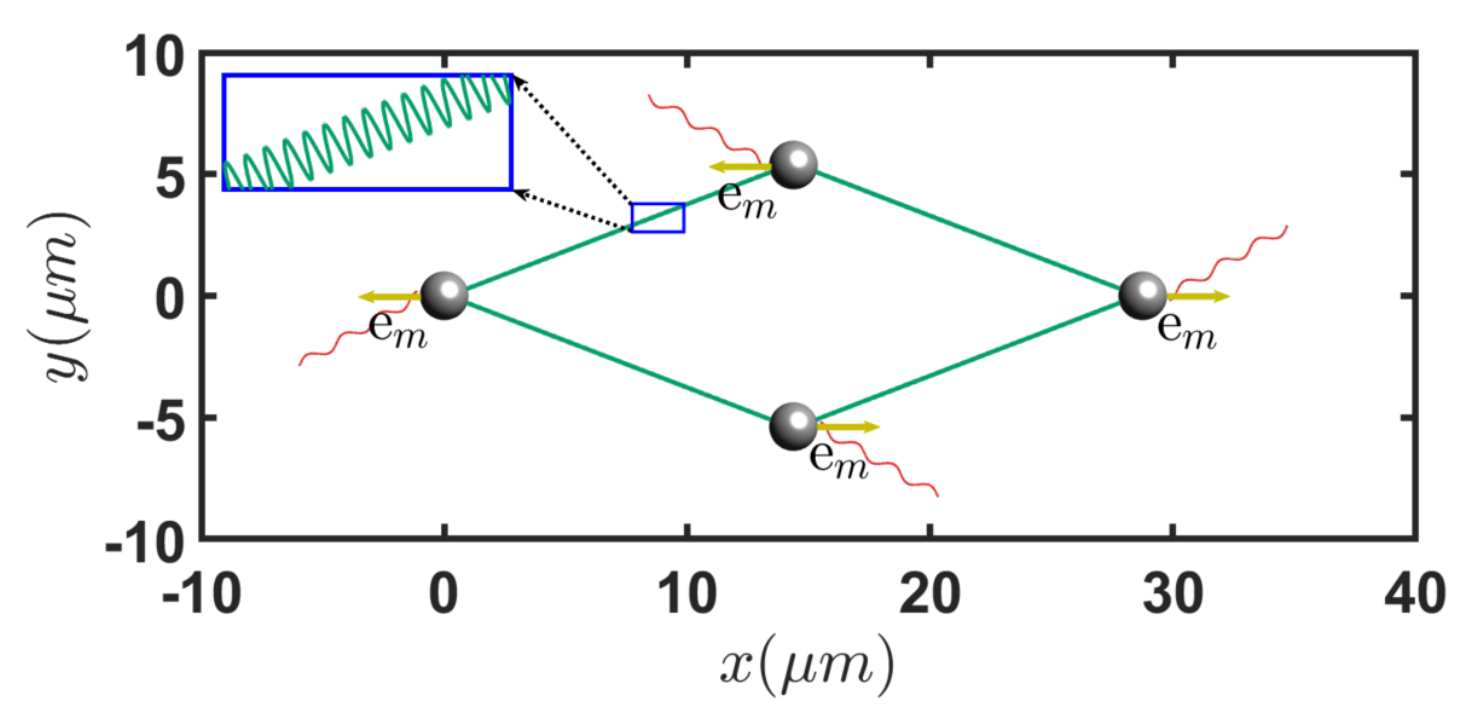}
\caption{Closed trajectory of a microswimmer's four-step locomotion involves alternating both the flagellar direction and the initial phase at each step, while adjusting the motor's rotation direction every two steps. Multimedia available online.}
\label{fig:fig4}
\end{figure}

From Fig.~\ref{fig:fig3}, it is clear that the direction of the flagellum is crucial for controlling the locomotion direction of the microswimmer. The angle between the flagellar axis and the forward direction, referred to as the yaw angle, plays a critical role in determining the navigation capabilities of the flagellum. To intuitively investigate the swimming trajectory of the microswimmer in the fluid, we simulate its locomotion in four steps, as illustrated in Fig.~\ref{fig:fig4} (Multimedia available online). In all four steps, the angle is set at $\psi=30^{\circ}$. In the first step, the motor rotates along the negative $x$-axis. In the second step, the flagellar axis is symmetrically reflected about the $x$-axis, and the initial phase of the flagellum is increased by half a cycle ($\pi$) to maintain the morphological symmetry of the microswimmer, while the motor's rotation direction remains unchanged. During the third step, the flagellar axis is symmetrically reflected about the $y$-axis, and the initial phase is again increased by half a cycle ($\pi$), with the motor's rotation direction now aligned along the positive $x$-axis. In the fourth step, the flagellar axis is symmetrically reflected about the $x$-axis once again, and the initial phase is increased by half a cycle ($\pi$), while the motor's rotation direction continues along the positive $x$-axis. Each step lasts for $3$ \si{s}, resulting in a total duration of 12 \si{s}, after which the microswimmer returns to its initial position. The enlarged portion indicates that the locomotion trajectory of the microswimmer follows a helical path. This simulation demonstrates that the forward direction of the microswimmer is governed by the flagellum, with changes in the forward direction corresponding to alterations in the flagellar direction~\cite{Lauga2016,Kuhn2017}. Therefore, navigation can be achieved by adjusting the flagellar direction, and the entire locomotion process follows a helical trajectory.

%%%%%%%%%%%%%%%%%%%%%%%%%%%%%%%%%%%%%%%%%%%%%%%%%%%%%%%%%%%%%%%%%%%%%%%%%%%%%%%%%%%%%%%%%%%%%%%%%%%%%%
\subsection{Hydrodynamic Interaction between Cell Body and Flagellum}
\begin{figure*}
\centering
\includegraphics[width=1.00\textwidth]{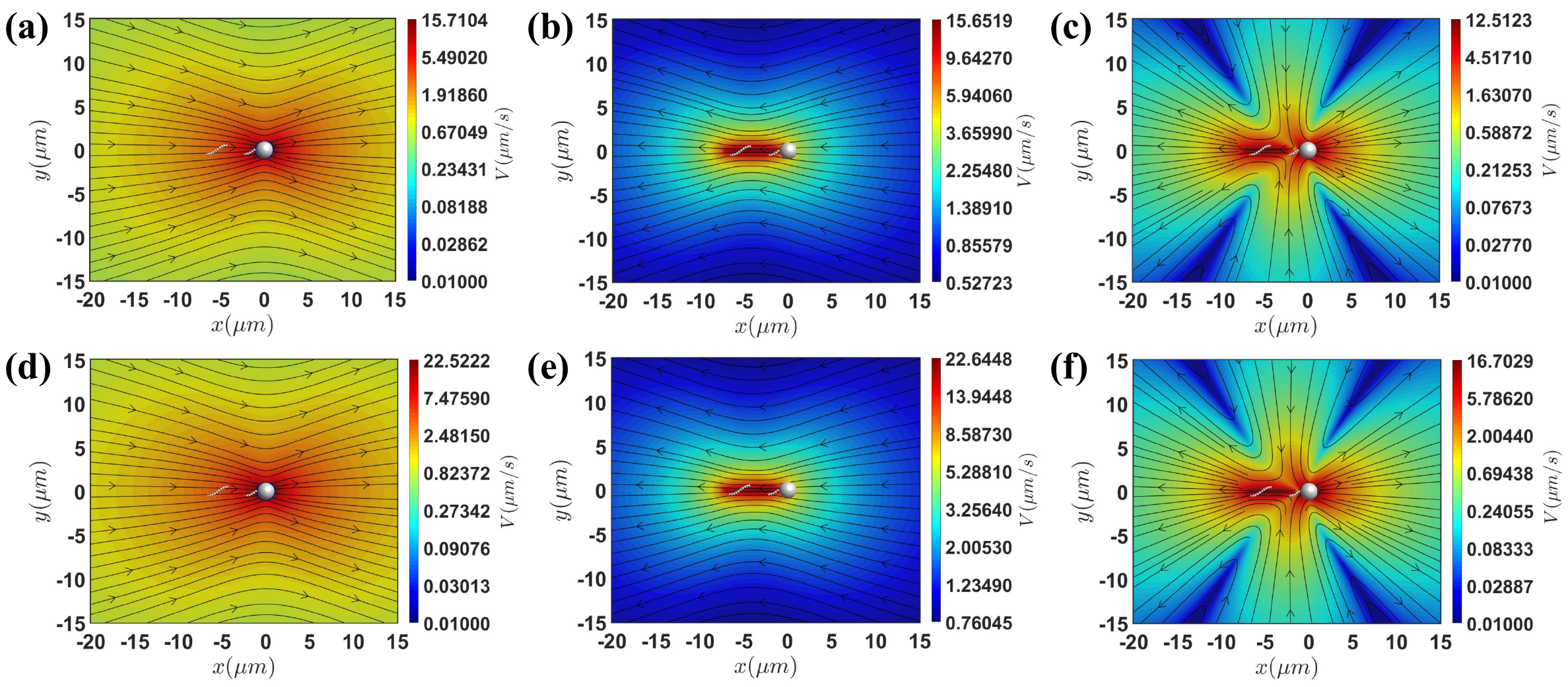}
\caption{(a), (b), (c) Flow fields generated by the cell body, flagellum, and the entire microswimmer, respectively, without considering the hydrodynamic interaction between the cell body and the flagellum. (d), (e), (f) Flow fields generated by the cell body, flagellum, and the entire microswimmer, respectively, while accounting for the hydrodynamic interaction between the cell body and the flagellum.}
\label{fig:fig5}
\end{figure*}

%%%%%%%%%%%%%%%%%%%%%%%%%%%%%%%%%%%%%%%%%%%%%%%%%%%%%%%%%%%%%%%%%%%%%%%%%%%%%%%%%%%%%%%%%%%%%%%%%%%%%%
The microswimmer model consists of a spherical cell body and a rigid helical flagellum. As the microswimmer moves through a fluid, the combined translational and rotational motions of both the cell body and the flagellum generate a complex flow field. To better visualize the flow fields produced by each component, we compute the flow field generated by the cell body, the flow field generated by the flagellum, and the overall flow field produced by the microswimmer separately, as depicted in Fig.~\ref{fig:fig5}. The flow around an isolated point force (Stokeslet) $\mathbf{F}$ located at the center of the sphere $\mathbf{X}_{0}$ is described by~\cite{Drescher2010,Tan2022}:
\begin{equation}
\begin{split}
&\textbf{V}_{st}(\mathbf{X};\mathbf{F},\mathbf{X}_{0})\\
&=\frac{1}{8\pi\mu}
\left[\frac{\mathbb{I}}{\vert\mathbf{X}-\mathbf{X}_{0}\vert}+\frac{\left(\mathbf{X}-\mathbf{X}_{0}\right)\otimes\left(\mathbf{X}-\mathbf{X}_{0}\right)}{\vert\mathbf{X}-\mathbf{X}_{0}\vert^{3}}\right]\cdot\mathbf{F}.
\label{eq:refname14}
\end{split}
\end{equation}
The flow field generated by a torque $\mathbf{T}$ (rotlet) located at a point $\mathbf{X}_{0}$ in an unbounded fluid is described by~\cite{Drescher2010,Tan2022}:
\begin{equation}
\begin{split}
\mathbf{V}_{rot}(\mathbf{X};\mathbf{T},\mathbf{X}_{0})=\frac{1}{8\pi\mu}\frac{\mathbf{T}\times\left(\mathbf{X}-\mathbf{X}_{0}\right)}{\vert\mathbf{X}-\mathbf{X}_{0}\vert^{3}}.
\label{eq:refname15}
\end{split}
\end{equation}
The total flow field induced by the force and torque exerted on the fluid by the sphere can be expressed as:
\begin{equation}
\begin{split}
\mathbf{V}_{tol}(\mathbf{X})=\mathbf{V}_{st}(\mathbf{X};\mathbf{F},\mathbf{X}_{0})+\mathbf{V}_{rot}(\mathbf{X};\mathbf{T},\mathbf{X}_{0}).
\label{eq:refname16}
\end{split}
\end{equation}
The flow field around the microswimmer is obtained by performing a phase average over its locomotion. The phase-averaged flow field in three dimensions is:
\begin{equation}
\begin{split}
\mathbf{V}(\mathbf{X})=\left\langle\sum_{i=1}^{N+1}\mathbf{V}_{tol}^{i}(\mathbf{X})\right\rangle.
\label{eq:refname17}
\end{split}
\end{equation}
where $\mathbf{V}_{tol}^{i}(\mathbf{X})$ represents the flow field generated by the $i$-th sphere of the microswimmer, consisting of $N$ spheres along the flagellum and a spherical cell body. Here, $\left\langle\cdot\right\rangle$ denotes the phase average.

As the microswimmer moves through the fluid, hydrodynamic interaction occurs between the cell body and the flagellum. To provide an intuitive understanding of how this interaction influences the swimming motion, we visualize the flow fields both with and without considering this interaction. Figs.~\ref{fig:fig5}(a), (b), and (c) depict the flow fields generated by the cell body, the flagellum, and the entire microswimmer, respectively, when hydrodynamic interaction is neglected. In contrast, Figs.~\ref{fig:fig5}(d), (e), and (f) present the flow fields when this interaction is considered. Notably, the flow field produced by the flagellum is directed opposite to the forward direction of the microswimmer, while the flow field generated by the cell body is aligned with its forward motion. Both flow fields experience a significant increase when the hydrodynamic interaction between the cell body and the flagellum is taken into account, indicating substantial enhancement of the thrust produced by the flagellum. Additionally, the overall flow field generated by the microswimmer is also significantly enhanced.

\begin{figure}
\centering
\includegraphics[width=0.50\textwidth]{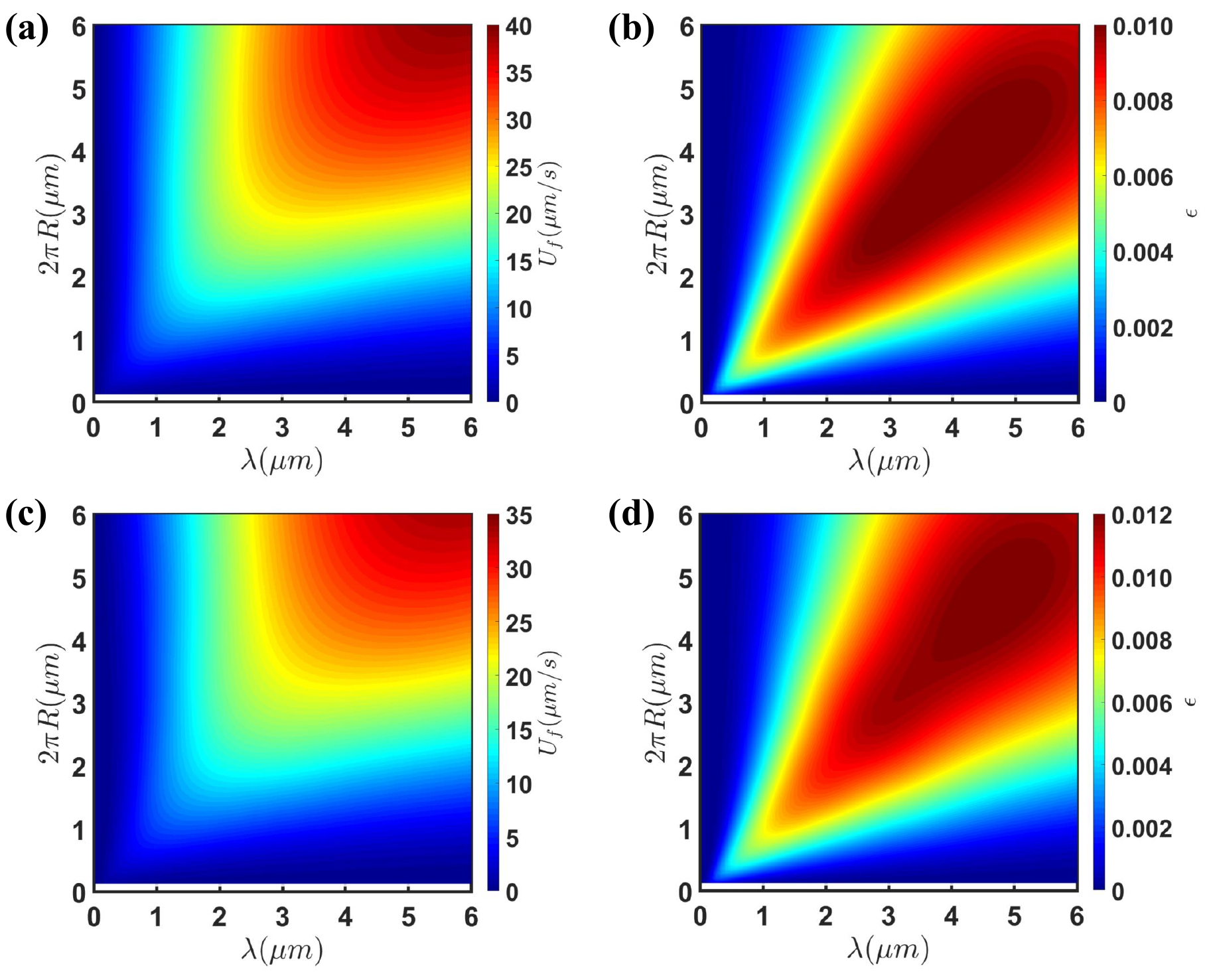}
\caption{(a), (c) Contours illustrating the forward speed of the microswimmer in the circumference-pitch plane. (a) corresponds to the scenario where the hydrodynamic interaction between the cell body and the flagellum is neglected, while (c) incorporates this interaction. (b), (d) Contours representing the swimming efficiency of the microswimmer in the circumference-pitch plane. (b) depicts the scenario without hydrodynamic interaction between the cell body and the flagellum, whereas (d) includes this interaction.}
\label{fig:fig6}
\end{figure}

Figures~\ref{fig:fig6}(a) and (c) present  the forward speed of the microswimmer in the circumference-pitch plane, where Fig.~\ref{fig:fig6}(a) represents the forward speed while neglecting the hydrodynamic interaction between the cell body and the flagellum, while Fig.~\ref{fig:fig6}(c) includes this interaction. Similarly, Figs.~\ref{fig:fig6}(b) and (d) illustrate the swimming efficiency of the microswimmer under the same conditions. In particular, when the hydrodynamic interaction between the cell body and the flagellum is considered, the forward speed of the microswimmer decreases, but the swimming efficiency increases. The opposing rotational directions of the cell body and the flagellum result in an ambient flow field where $\Omega_{x}^{\infty}<0$. According to the relationship $\mathcal{R}_{xx}^{rr}(\Omega_{x}-\Omega_{x}^{\infty})=T_{x}$, the torque exerted on the cell body by the motor increases when this hydrodynamic interaction is taken into account. As shown in Figs.~\ref{fig:fig5}(b) and (e), the ambient translational flow field generated by the flagellum also satisfies $U_{x}^{\infty}<0$. Thus, based on the relationship $\mathcal{R}_{xx}^{tt}(U_{x}-U_{x}^{\infty})=F_{x}$, the propulsive force exerted by the flagellum on the microswimmer increases when considering this hydrodynamic interaction. In this context, the propulsive force with hydrodynamic interaction is approximately $1.5$ times greater than that when this interaction is neglected. According to the swimming efficiency formula Eq.~\ref{eq:refname7}, the variation in swimming efficiency results from the competition between the propulsive force and the torque.

%%%%%%%%%%%%%%%%%%%%%%%%%%%%%%%%%%%%%%%%%%%%%%%%%%%%%%%%%%%%%%%%%%%%%%%%%%%%%%%%%%%%%%%%%%%%%%%%%%%%%%
\subsection{Optimal Morphology of Flagella}
%%%%%%%%%%%%%%%%%%%%%%%%%%%%%%%%%%%%%%%%%%%%%%%%%%%%%%%%%%%%%%%%%%%%%%%%%%%%%%%%%%%%%%%%%%%%%%%%%%%%%%
The trajectories of the microswimmers, illustrated in Fig.~\ref{fig:fig2} and Fig.~\ref{fig:fig4}, along with the forward speed and swimming efficiency depicted on the circumference-pitch plane in Fig.~\ref{fig:fig6}, demonstrate that flagellar morphology has a significant impact on both forward speed and swimming efficiency. To systematically explore how the morphological parameters, including filament radius $a$, helix radius $R$, pitch angle $\theta$, and contour length $\Lambda$, affect the forward speed and swimming efficiency of the microswimmer, we analyze these metrics on the $\Lambda-2\theta/\pi$ plane and the $a-2\theta/\pi$ plane.

\begin{figure}
\centering
\includegraphics[width=0.50\textwidth]{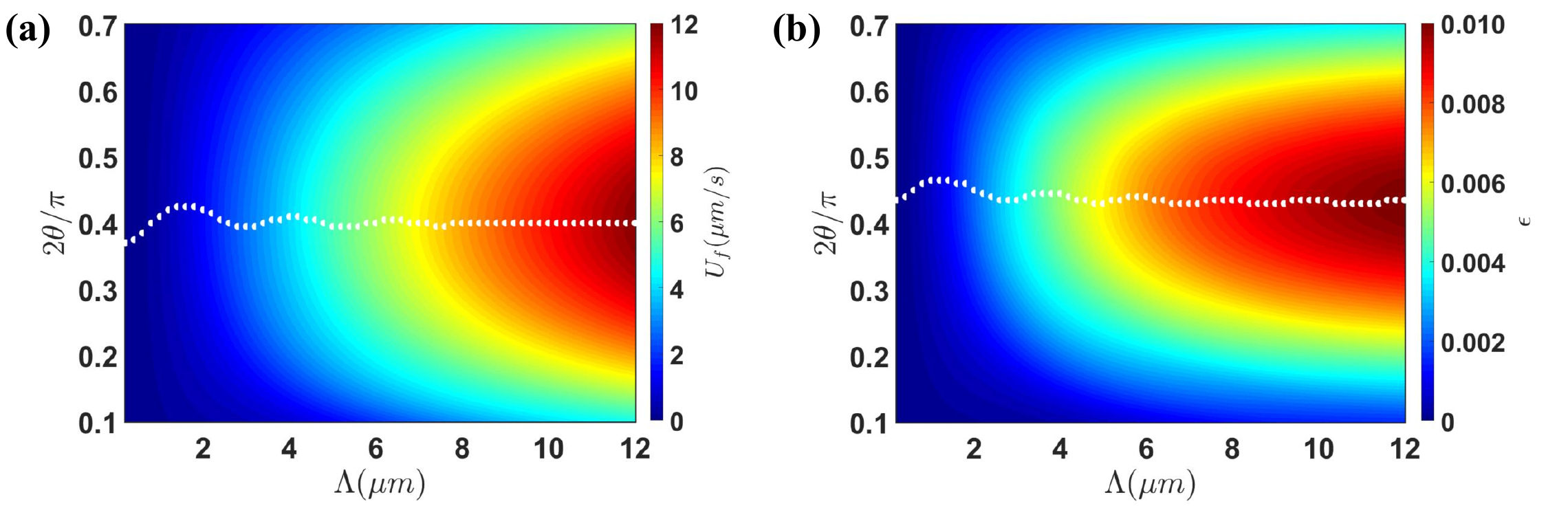}
\caption{(a) Contours of forward speed in the contour length-pitch angle plane for a freely swimming microswimmer with a helix radius of $R=0.2$ \si{\mu m}. (b) Contours of swimming efficiency in the contour length-pitch angle plane for the same microswimmer.}
\label{fig:fig7}
\end{figure}

Firstly, we assess the impact of flagellar contour length on the forward speed and swimming efficiency of the microswimmer. Using a helix radius of $R=0.2$ \si{\mu m}, we present the contours of forward speed and swimming efficiency for a freely swimming microswimmer in Fig.~\ref{fig:fig7}. The analysis reveals that for varying contour lengths, there is an optimal pitch angle, approximately $\pi/5$, which maximizes either forward speed or swimming efficiency. Additionally, the optimal pitch angles for contour lengths of $\Lambda=2$ \si{\mu m} and $\Lambda=4$ \si{\mu m} are slightly higher than those for the other contour lengths.

\begin{figure}
\centering
\includegraphics[width=0.50\textwidth]{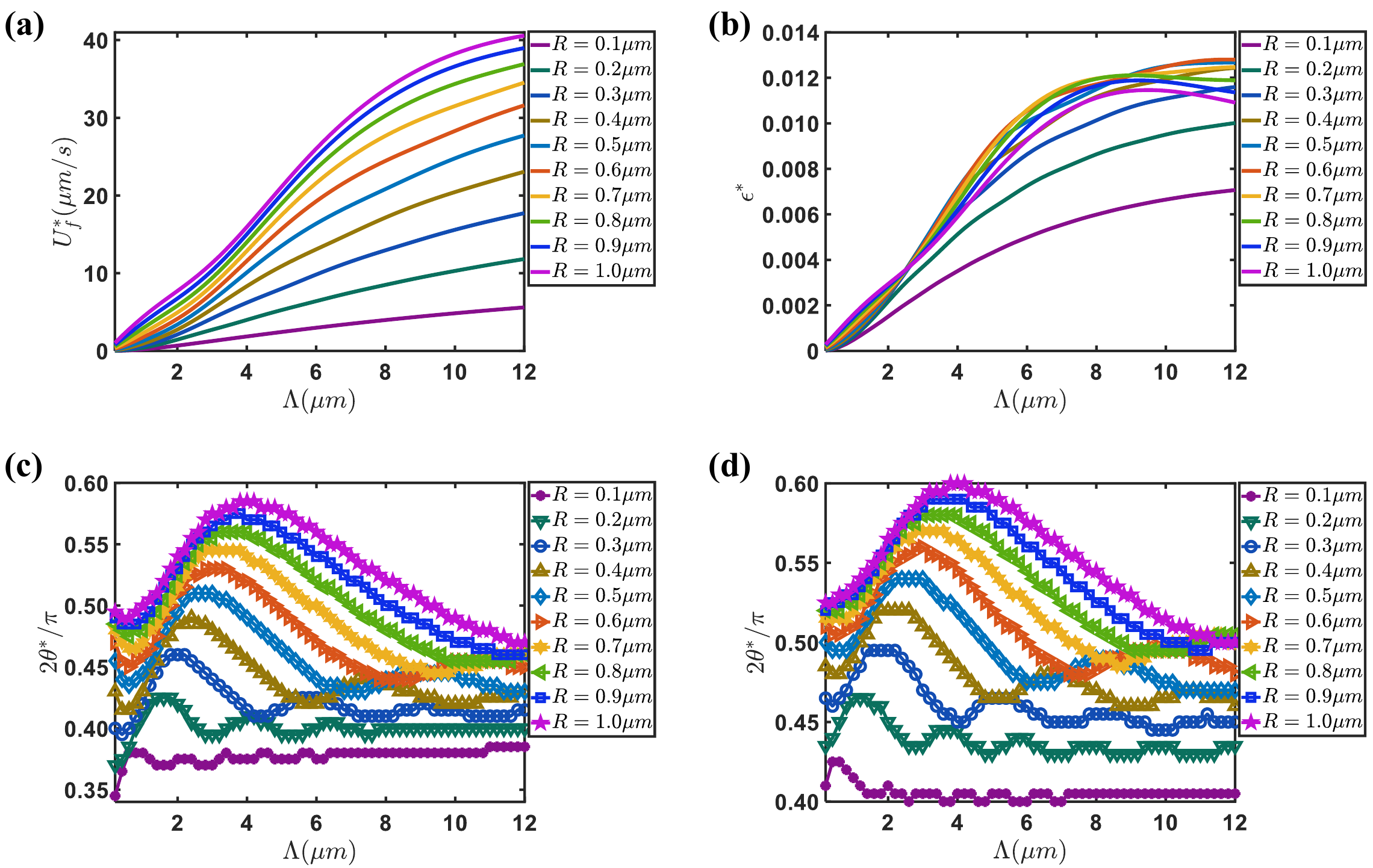}
\caption{(a) Variation of maximum forward speed as a function of contour length for various flagellar helix radii $R$. (b) Variation of maximum swimming efficiency as a function of contour length for different flagellar helix radii $R$. (c) Optimal pitch angle corresponding to the maximum forward speed. (d) Optimal pitch angle corresponding to the maximum swimming efficiency.}
\label{fig:fig8}
\end{figure}

We examine various helix radii, $R$, to calculate the maximum forward speed and maximum swimming efficiency across different contour lengths and pitch angles, along with their corresponding optimal pitch angles, as illustrated in Fig.~\ref{fig:fig8}. As shown in Fig.~\ref{fig:fig8}(a), the maximum forward speed increases with both the helix radius and contour length. However, once the helix radius exceeds $0.3$ \si{\mu m}, the maximum efficiency reaches a plateau and does not improve further, as depicted in Fig.~\ref{fig:fig8}(b). Therefore, for a cell body radius of $R_{b}=1$ \si{\mu m}, the optimal helix radius is approximately in the range of $0.2 \leq R \leq 0.3$ \si{\mu m}, and the corresponding optimal pitch angles range from $30^{\circ}$ to $45^{\circ}$. Moreover, for the majority of flagellar contour lengths, it is observed that the optimal pitch angle increases with the helix radius, as shown in Figs.~\ref{fig:fig8}(c) and (d).

\begin{figure}
\centering
\includegraphics[width=0.50\textwidth]{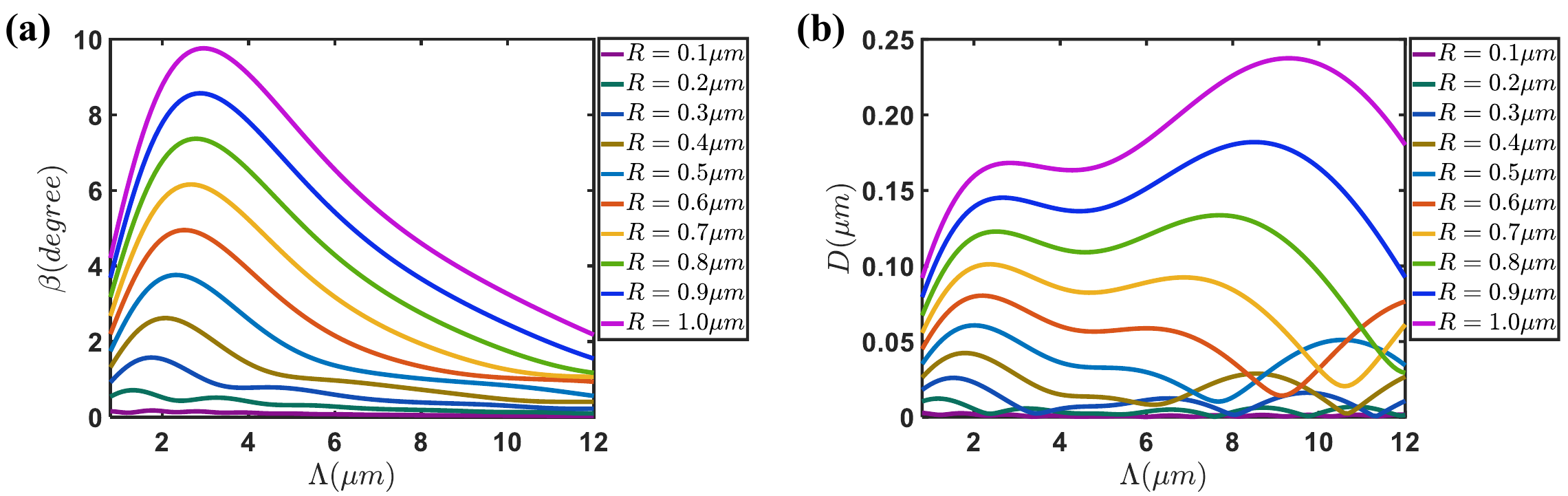}
\caption{(a) Variation of the yaw angle with contour length $\Lambda$ for different helix radii $R$. (b) Variation of the helical diameter of the microswimmer's trajectory with contour length $\Lambda$ for different helix radii $R$.}
\label{fig:fig9}
\end{figure}

Microswimmers need to capture nutrients to survive in fluid environments~\cite{Guasto2012,Yawata2014}. As shown in Figs.~\ref{fig:fig2} and \ref{fig:fig4}, the trajectories of microswimmers exhibit a circular helix, allowing them to navigate by adjusting the orientation of their flagella. In addition to considering forward speed and swimming efficiency, it is crucial to evaluate their directional capabilities for survival in these environments. To quantitatively characterize this capability, we introduce the yaw angle $\beta$, defined as the angle between the flagellar axis and the direction of forward locomotion. Fig.~\ref{fig:fig9}(a) demonstrates that a smaller helix radius is associated with a smaller yaw angle, indicating enhanced directionality. Furthermore, Fig.~\ref{fig:fig8} shows that when the helix radius $R\le3$ \si{\mu m}, both forward speed and swimming efficiency decrease as the helix radius decreases. Analyzing Fig.~\ref{fig:fig9}(a) further confirms that a smaller flagellum helix radius correlates with stronger directional control. Additionally, Fig.~\ref{fig:fig9}(b) indicates that when $R<0.3$ \si{\mu m}, the helical diameter of the trajectory is less than $26$ \si{nm}, which is comparable to the size of the nutrients necessary for the microswimmer's survival. As the helical radius of the microswimmer's trajectory decreases, its efficiency in capturing nutrients increases. By integrating factors such as forward speed, swimming efficiency, yaw angle, and diameter of the helical trajectory, we conclude that the optimal range for the flagellar helix radius is $0.2 \leq R \leq 0.3$ \si{\mu m}.

\begin{figure}
\centering
\includegraphics[width=0.50\textwidth]{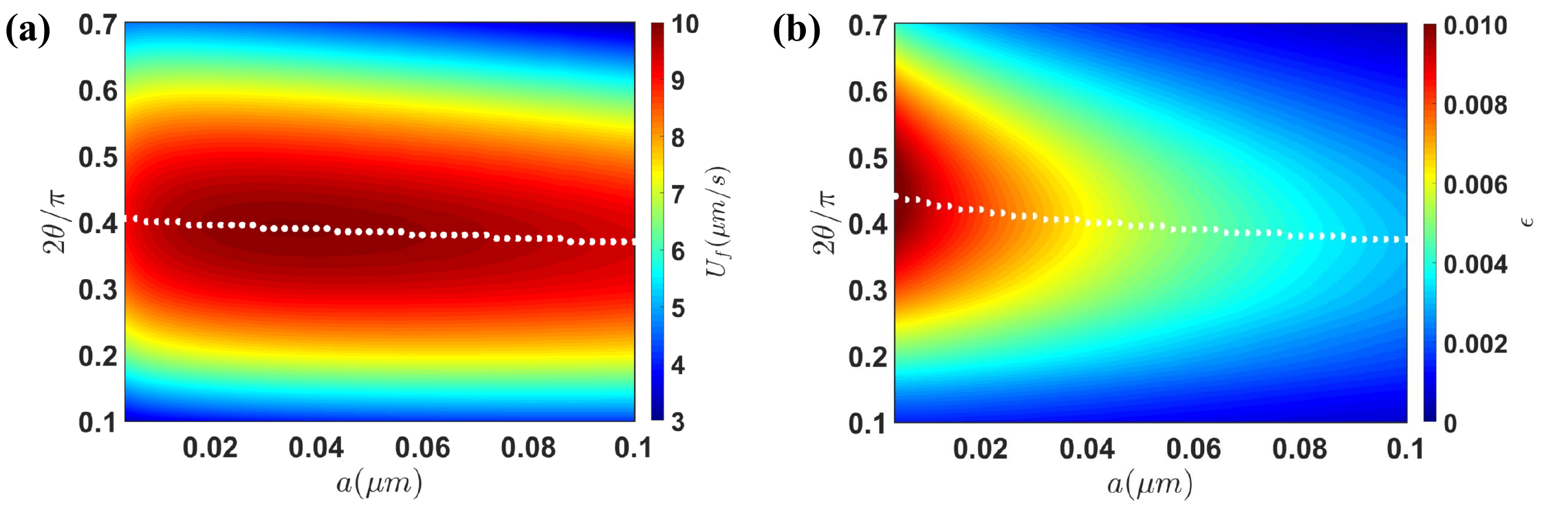}
\caption{(a) Forward speed contours in the filament radius-pitch angle plane for a freely swimming microswimmer with a contour length of $\Lambda=9$ \si{\mu m}. (b) Swimming efficiency contours in the filament radius-pitch angle plane for the same microswimmer.}
\label{fig:fig10}
\end{figure}

To examine the effects of filament radius and pitch angle on microswimmers, we select a flagellum with a contour length of $\Lambda=10$ \si{\mu m} and a helix radius of $R=0.2$ \si{\mu m}. By systematically varying the filament radius and pitch angle, we simulate the forward speed and swimming efficiency of the microswimmer, as illustrated in Fig.~\ref{fig:fig10}. The results reveal that for different filament radii $a$, there exists an optimal pitch angle $\theta$ that maximizes either the forward speed or swimming efficiency. Moreover, as shown in Fig.~\ref{fig:fig10}(a), an optimal filament radius can also be identified that maximizes the forward speed. While a smaller filament radius enhances swimming efficiency (Fig.~\ref{fig:fig10}(b)), the construction constraints of the polymer used for the flagellum require that the filament radius remains above the nanometer scale.

\begin{figure}
\centering
\includegraphics[width=0.50\textwidth]{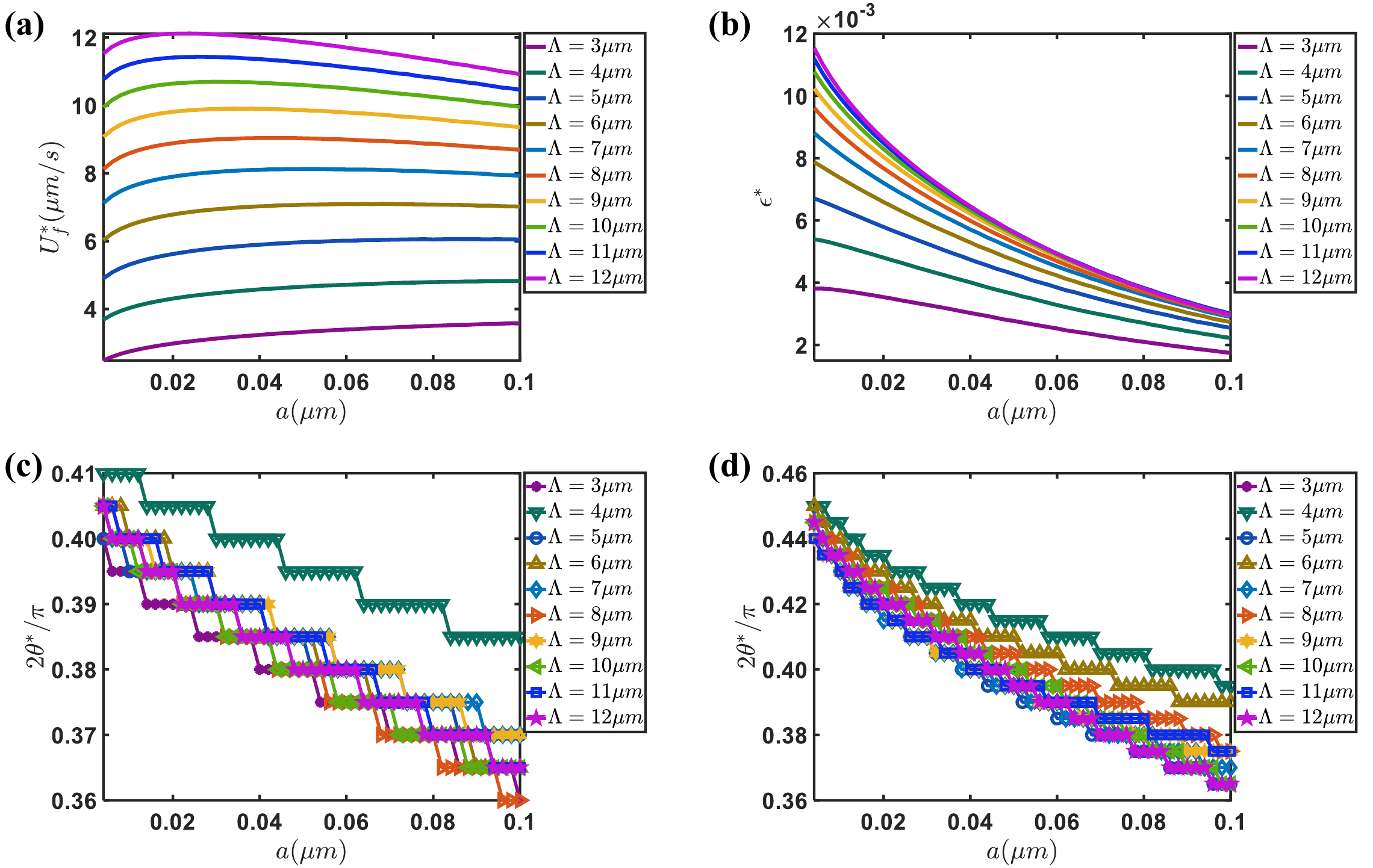}
\caption{(a) Maximum forward speed as a function of filament radius for various flagellar contour lengths. (b) Maximum swimming efficiency as a function of filament radius for different flagellar contour lengths. (c) Optimal pitch angle corresponding to the maximum forward speed. (d) Optimal pitch angle corresponding to the maximum swimming efficiency.}
\label{fig:fig11}
\end{figure}

Figure~\ref{fig:fig11} demonstrates the existence of an optimal pitch angle for the flagellum that maximizes either the forward speed or swimming efficiency of the microswimmer. We present contours of forward speed and swimming efficiency with respect to various flagellar contour lengths, from which we extract the maximum forward speed, maximum swimming efficiency, and their corresponding optimal pitch angles, as shown in Fig.~\ref{fig:fig11}. From Fig.~\ref{fig:fig11}(a), it is clear that the filament radius has a minimal impact on the maximum forward speed of the microswimmer. However, maximum swimming efficiency significantly decreases as the filament radii increase, as illustrated in Fig.~\ref{fig:fig11}(b). This suggests that while a smaller filament radius can enhance swimming efficiency, when the contour length of the flagellum exceeds $9$ \si{\mu m}, the improvement in swimming efficiency becomes negligible. Therefore, longer flagellar contour lengths do not necessarily yield better performance for the microswimmer.

When the filament radius is less than $0.07$ \si{\mu m}, the pitch angle for maximum forward speed is slightly lower than that for maximum swimming efficiency, with a difference not exceeding $4.5^{\circ}$. This minor variation allows the microswimmer to make slight adjustments to the pitch angle, facilitating a transition between flagellar morphologies optimized for either maximum forward speed or maximum swimming efficiency. This adaptability is essential for enhancing the survival capability of the microswimmer. This also explains why the Young's modulus of most flagella is relatively high, as a low Young's modulus would be detrimental to bacterial survival~\cite{Darnton2007,Son2013,Vogel2013}. From Figs.~\ref{fig:fig11}(c) and (d), it is evident that when the filament radius falls within the range $0.01 \leqslant a \leqslant 0.02$ \si{\mu m}, the pitch angles corresponding to both maximum forward speed and optimal swimming efficiency are approximately $30^{\circ} \leqslant \theta \leqslant 45^{\circ}$. This pitch angle range aligns with the typical morphology of the flagellum~\cite{Koyasu1984,Fujii2008,Spagnolie2011}.

%%%%%%%%%%%%%%%%%%%%%%%%%%%%%%%%%%%%%%%%%%%%%%%%%%%%%%%%%%%%%%%%%%%%%%%%%%%%%%%%%%%%%%%%%%%%%%%%%%%%%%
\section{Summary and Conclusions}
%%%%%%%%%%%%%%%%%%%%%%%%%%%%%%%%%%%%%%%%%%%%%%%%%%%%%%%%%%%%%%%%%%%%%%%%%%%%%%%%%%%%%%%%%%%%%%%%%%%%%%
The helical flagella of microswimmers exhibit axial asymmetry, resulting in uneven force distributions during rotation in a fluid. This asymmetry induces precessional motion as the microswimmers swim, with the precession angle closely linked to the morphology of the flagella. Consequently, the motion trajectory of the microswimmer resembles a circular helix with a distinct yaw angle between its forward direction and the flagellar axis. By adjusting the orientation of the flagella, the microswimmer can effectively change its forward direction, enabling precise navigation.

During swimming, the flagella generate a flow field that opposes the forward motion of the microswimmer, while the cell body produces a flow field in the same direction as the forward movement. Such hydrodynamic interaction between the cell body and the flagella is crucial. Our findings indicate that considering this interaction significantly increases the propulsive force generated by the flagella. Although this interaction may reduce the forward speed, it enhances swimming efficiency.

The morphology of flagella is essential for determining the forward velocity, swimming efficiency, and directional stability of microswimmers. Research indicates that an optimal flagellar morphology exists that maximizes either forward velocity or swimming efficiency. The optimal contour length of the flagellum is around $10$ \si{\mu m}. Beyond this length, additional increases in contour length do not significantly enhance the swimming efficiency of the microswimmer. Although the radius of the filaments has a relatively minor effect on forward velocity, reducing the filament radius significantly enhances swimming efficiency. The optimal helix radius for microswimmers is approximately $0.2$ to $0.3$ \si{\mu m}. Beyond this range, increasing the helix radius does not further improve swimming efficiency; rather, it increases the yaw angle and the diameter of the helical trajectory, adversely affecting the directional stability. At a helix radius of $R=0.2$ \si{\mu m}, the optimal pitch angle for maximum forward speed is slightly lower, by no more than $4.5^{\circ}$, than that for maximum swimming efficiency. This flexibility enables microswimmers to transition between maximum forward speed and maximum swimming efficiency by fine-tuning the pitch angle, with the optimal pitch angle range being approximately $30^{\circ}$ to $45^{\circ}$. Therefore, the flagellum only needs to make minor adjustments to its morphology to adapt to different living environments, which is why its Young's modulus is relatively high.

%%%%%%%%%%%%%%%%%%%%%%%%%%%%%%%%%%%%%%%%%%%%%%%%%%%%%%%%%%%%%%%%%%%%%%%%%%%%%%%%%%%%%%%%%%%%%%%%%%%%%%
\begin{acknowledgments}
We acknowledge computational support from Beijing Computational Science Research Center. This work is supported by the National Natural Science Foundation of China(NSFC) umder Grant No. U2230402.
\end{acknowledgments}

%%%%%%%%%%%%%%%%%%%%%%%%%%%%%%%%%%%%%%%%%%%%%%%%%%%%%%%%%%%%%%%%%%%%%%%%%%%%%%%%%%%%%%%%%%%%%%%%%%%%%%
%%%%%%%%%%%%%%%%%%%%%%%%%%%%%%%%%%%%%%%%%%%%%%%%%%%%%%%%%%%%%%%%%%%%%%%%%%%%%%%%%%%%%%%%%%%%%%%%%%%%%%
\nocite{*}
\bibliography{aipsamp}% Produces the bibliography via BibTeX.

\end{document}